 \documentstyle [epsf,multicol,aps,prl,epsfig] {revtex} 

\newcommand{\be}{\begin{equation}}
\newcommand{\ee}{\end{equation}}
\newcommand{\bea}{\begin{eqnarray}}
\newcommand{\eea}{\end{eqnarray}}
\newcommand{\bsa}{\begin{subeqnarray}}
\newcommand{\esa}{\end{subeqnarray}}

\begin{document}
\title{Frictional shear cracks}
\author{ Efim A. Brener $^{1)}$ and V.I. Marchenko $^{2)}$ }
\address{1) Institut f\"ur Festk\"orperforschung, Forschungszentrum
J\"ulich,\\
D-52425 J\"ulich, Germany}
\address{2) P.L. Kapitza Institute for Physical Problems, RAS,\\
117334, Kosygin str. 2, Moscow, Russia}
\author{(\today)}
\author{\parbox{397pt}
{\vglue 0.3cm \small We discuss  crack propagation along the
interface between two dissimilar materials. The crack edge separates
two states of the interface, ``stick'' and ``slip''. In the slip
region we assume  that the shear stress is proportional to the
sliding velocity, i.e. the linear viscous friction law. In this picture
the static friction appears  as the Griffith threshold for
crack propagation. We calculate the crack velocity as a function
of the applied shear stress and find that the main dissipation
comes from the macroscopic region and is mainly due to the
friction at the interface. The relevance of our results to 
recent experiments, Baumberger {\it et al}., Phys. Rev. Lett.
{\bf 88}, 075509 (2002),
is discussed.}}
\author{\parbox{397pt}{\vglue 0.3cm \small
}}
\maketitle
\begin{multicols}{2}
\narrowtext A few recent experimental observations of Rubio and
Galeano \cite{Rubio94}, and Baumberger, Caroli, and Ronsin
\cite{Baumberger2002}, on the frictional motion of sheared gels
sliding along a glass surface indicate the existence of self-healing
pulses and inhomogeneous modes of sliding \cite{Persson2000}. 
A regime of periodic
stick slip has been observed in a limited range of small
shearing rates \cite{Baumberger2002}. It bifurcates towards
stationary sliding at some critical driving velocity. The slip
pulses traverse the sample with a velocity much larger than the
driving velocity but still much smaller than the  speed of sound.

Slip pulses in gels seem to be very different from Schallamach
waves and ``brittle'' pulses studied by Gerde and Marder
\cite{Gerde01} since no observable interface separation occurs.
In this respect, they are more comparable with
self-healing cracks suggested by Heaton \cite{Heaton90} in the
context of seismic events.

Recent investigations (see, for example, \cite{Ranjith01} and
references therein) point towards an essential importance of the
underlying friction law in the slip state. It has been proved that
the simple Coulomb friction leads to the so-called
``ill-posedness'' of the linear stability problem while discussing
small nonhomogeneous perturbations of the stress and strain fields
in a sliding mode \cite{Ranjith01}. Moreover, Caroli
\cite{Caroli2000} has shown that the existence of slow, periodic
slip pulses is incompatible with Coulomb friction law.

In this letter we discuss  crack propagation along the interface
between two dissimilar materials. The crack edge separates two states
of the interface, ``stick'' and ``slip''. We assume that the
interface is flat with a strong adhesion contact. In principle, we
could allow for small wavelength surface roughness, but in this
case we consider  length scales larger than the longest
wavelength component. In the presence of roughness, the assumption of 
strong adhesion and full contact at the interface presumably
is only reasonable  for  ``soft''
materials with relatively small shear modulus. Gels are clearly
materials of this sort.

In the slip region we assume a simple linear viscous  friction law,
namely, the shear stress is proportional to the sliding velocity.
This from the theoretical point of view strongly motivated law 
is  usually  not discussed in literature since it
does not lead to the so-called static friction phenomenon observed
experimentally. However, we will see that in our description  static 
friction appears in a natural way  as the usual Griffith
threshold for crack propagation. The important point is that
before the system goes  into a sliding mode the slip pulse should
traverse the sample. This requires finite shear stress since the
stick state of the interface is energetically more favorable.

With the linear viscous friction law we find conditions for crack
propagation and calculate the crack velocity as a function of an
applied shear stress. We find that the main dissipation comes from
the macroscopic region and  is due to the friction at the interface.
This situation is very different from the usual crack propagation
where the main dissipation is localized in the microscopic tip
region.

We also shortly discuss frictional shear cracks inside homogeneous
materials. The point here is that in mode II (and in mode III)
cracks there is no macroscopic opening. If  two surfaces remain in
contact, the standard boundary conditions, namely the vanishing of 
normal and shear stresses on the crack surfaces, are not
theoretically motivated. The relative sliding velocity of the two
surfaces should lead to   nonzero shear stresses.
Finally we discuss the relevance of our results to the experimental
observations  \cite{Baumberger2002}.

Consider an elastic solid sliding on a flat rigid substrate.
Assume that the elastic solid occupies the space $H>y>0$, and let
$(x,y,z)$ be a coordinate system with the  plane $y=0$ corresponding to
the surface  of the solid, see Fig.1. We discuss the plane strain
situation with $u_z=0$, where $\bf u$ is the displacement vector. We
assume that the interface can be in two states: "stick" and
"slip". The boundary between these two states is described by the
crack edge which moves with a  velocity $V_{tip}$ in the  $x$
direction. In the stick region the displacements are continuous
and since we assume a rigid substrate, the boundary conditions
are: $u_x=u_y=0$ for $x-V_{tip}t>0$ and $y=0$. In the slip region
we assume that the two solids (for all times) are in contact,
$u_y=0$ for $x-V_{tip}t<0$ and $y=0$, while we allow for a finite
relative sliding velocity $\dot u_x$. This sliding velocity leads
to   frictional shear stress at the interface
where we assume a linear viscous friction law:
\be
\sigma _{xy}=\alpha \dot u_x,
\ee
with $\alpha$ being the viscous friction coefficient.
\begin{figure}[h]
\epsfig{figure=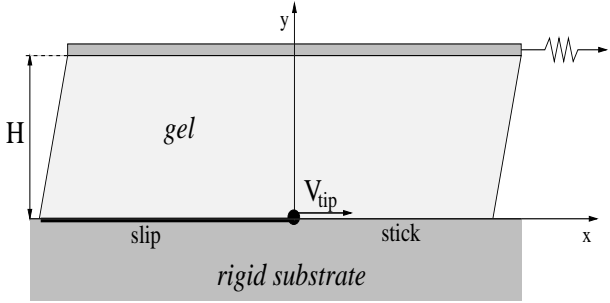, width=8cm,height=4cm,clip=}
\caption{An elastic body sliding on a rigid substrate}
\hfill
\end{figure}
It is reasonable to assume that the interface energy in
the stick phase is smaller than the interface energy in the slip
phase since the adhesion contact in the stick region is stronger.
Let us denote this energy difference by $\gamma$. It is clear that
without external loading the stick phase is energetically
favorable and a finite shear stress is required to get the interface
into the slip state. Let us assume that far ahead of the crack tip,
the solid is  homogeneously strained with $u_{xy}^{\infty}$ and
stressed with $\sigma_{xy}=2\mu u_{xy}^{\infty}$ where $\mu$ is
the shear modulus. The strain energy is $\mu
(u_{xy}^{\infty})^2H$. Far behind the crack tip where the stress is
relaxed, only the interface energy $\gamma$ remains. The slip state
will be realized only if $\Delta=  \mu (u_{xy}^{\infty})^2H/\gamma
>1$. In this case the crack should propagate in the positive $x$
direction. Otherwise, the crack would propagate with negative
velocity and the stick phase will be restored. Condition
$\Delta=1$ is nothing but the usual Griffith threshold for crack
propagation. On the other hand, in the context of the friction
problem, this condition may be interpreted as a static friction
threshold: a finite shear loading is required to get the system into the
sliding mode.

If the whole interface is in the slip state, a steady-state
motion of the elastic body is possible with a velocity
\be
V=2Su_{xy},
\ee
where $S=\mu/\alpha$ is the velocity scale given
by the friction law. We note that this homogeneous sliding mode is
linearly stable for any velocity with respect to small
nonhomogeneous perturbations of the stress and strain fields
localized in the surface region. In this respect the viscous
friction law is very different from Coulomb friction which leads
to a linear instability and ill-posedness of the problem as it
has been intensively discussed in  literature \cite{Ranjith01}.

On the other hand, the homogeneous sliding mode may be 
unstable against the resticking pulse 
(nonlinear ``healing instability'') if the
corresponding value of $\Delta <1$. Since in this case, the strain 
which defines the value of $\Delta$  is related to the
steady-state sliding velocity by Eq.(2), we find that the
homogeneous sliding is stable against the healing instability only
above the critical sliding velocity
\be
V_c=2S(\gamma/\mu H)^{1/2}.
\ee

Now let us turn to the calculation of the crack tip velocity
$V_{tip}$ as a function of the dimensionless driving force
$\Delta$. The strategy is as follows: we solve the elastic problem
in the vicinity of the crack tip and then calculate the energy
flux into the crack tip and the dissipation due to the friction at the
interface. Finally, using
the energy balance, we find the crack velocity.

Let us start from some qualitative estimates. Assume that, as in
the usual crack problem, the singular behavior of the displacement
vector is given by a square-root singularity in the vicinity of
the crack tip. Then the dissipation rate at the interface \be J_d=
\alpha \int (\dot u_x)^2dx \ee diverges logarithmically. This is
already quite a remarkable observation. In the usual crack problem
the main dissipation comes from the close vicinity of the tip and
often requires the introduction of microscopic models. Here we
have the chance to avoid such a detailed microscopic description by
using some microscopic length scale as a cutoff which enters only
the logarithm in the final result. Note that  dissipation
due to the bulk viscoelasticity effectively only leads  to  tip
dissipation on microscales. Indeed, the viscoelasticity gives
corrections to the stress tensor of the form $\eta \dot
u_{ik}$. The dissipation rate due to this effect diverges
strongly at small distances  (as $1/r$) 
and correspondingly decays
at  macroscopic distances. Thus, this effect can be incorporated
into the tip dissipation.

Now we solve the elastic problem more accurately, while still using a
quasistatic approximation for the moment. The generalization to
the full elastodynamic description is straightforward and will be
given below. In the co-moving frame of reference and in the 
vicinity  of the crack tip,  the displacement
field for the static elasticity and the boundary conditions
formulated above is 
$$ u_x=ARe\left[ yz^{\lambda -1}-i\left(
3-4\nu \right) z^\lambda/\lambda \right],
$$
\be
u_y=ARe\left[ i y z^{\lambda -1}\right]. \ee Here $z$ is a complex
coordinate $z=x+iy$,  $\nu $ is the Poisson ratio, and
$A$ is a real amplitude. The spectrum of $\lambda $ is 
purely real and given by the following equation:
\be
\exp \left( i2\pi \lambda \right) =-\frac{1+i\pi
\varepsilon /2}{1-i\pi\varepsilon /2},
\ee
with
\be
\varepsilon =\frac{1}{2\pi }\frac{3-4\nu } {1-\nu
}\frac{V_{tip}}{S}.
\ee
In the limit of  small values of 
$\left|\varepsilon \right| $ for the leading crack
displacement component, we have $2\lambda \approx 1+\varepsilon $.

Having defined the displacement field we can calculate the energy flux
 $J_i=\sigma _{ik}\dot u_k$ and the local energy release into a small
semicircular region with some microscopic radius $a$ around the
crack tip,
\be
J_0=2\pi \mu (3-4\nu )(1-\nu )V_{tip}A^2 a^\varepsilon. 
\ee 
The
dissipation rate due to the interface friction with  exclusion
of the small region of  size  $a$ close to the tip  is given by Eq.
(4):
\be
J_d= \alpha  (3-4\nu )^2 A^2 V_{tip}^2 \varepsilon^{-1} \left(
\widetilde{H} ^\varepsilon -a^\varepsilon
\right)=J_0\left[(\widetilde{H}/a)^\varepsilon-1\right]. 
\ee 
Here
$\widetilde{H}=f(\nu )H$, with  $f$ being an undetermined yet
function of the
Poisson ratio; $H$ is a thickness of the
sample. This function can be found by solving
the elastic problem for  given geometry 
and all boundary conditions. As we will see the tip velocity does
not crucially depends  on the actual value of the factor $f$, which is
of the order of unity. 

On the other hand the local energy release into the crack tip
$J_0$ must compensate the surface energy difference: $J_0=\gamma
V_{tip}$. Note, that here we have neglected the dissipation at the
tip in comparison with the energy release $\gamma V_{tip}$, which
is reasonable  at small tip velocities  compared to
the velocity of sound.  Finally, using the global energy conservation
law,
\be
J_{\infty
}=J_{0}+J_{d}=\mu
\left( u_{xy}^{\infty }\right)^2 HV_{tip},
\ee
 we find
\be
\varepsilon=\frac{\ln\Delta}{\ln(\widetilde{H}/a)}\approx
\frac{\ln\Delta}{\ln(H/a)}.
\ee 
Since $\varepsilon$ is given by Eq. (7), 
this result is a compact
representation of the crack velocity as a function of the driving
force $\Delta =\mu \left( u_{xy}^{\infty }\right)^2 H/ \gamma $. Note
that $\Delta =1$ corresponds to Griffith equilibrium. 
Eq. (11)  is valid for small $|\varepsilon |$. The explicit
expression for the crack velocity reads
\be
V_{tip}=2\pi \frac{1-\nu }{3-4\nu }\frac{\ln \Delta}{\ln(H/a)}S.
\ee 
In the case of small $\Delta - 1$ we obtain
\be
V_{tip}\approx 2\pi \frac{1-\nu }{3-4\nu }\frac{\Delta -1} {\ln (H/a)}S.
\ee 
This result corresponds to a small dissipation rate compared
to the total energy flux, $J_d\ll J_{\infty}$, and can also be obtained
 using  perturbation theory: we solve the elastic
problem neglecting friction ($\sigma_{xy}=0$ at the interface) and
then calculate the dissipation rate (4) using this solution as zero order 
 displacement field.

Up to now we have used the static approximation. The tip velocity
should be small  compared to the  sound velocity. 
The elastodynamic generalization is straightforward.
Using the standard approach to the singular solutions near the 
crack tip \cite{Freund}, we find the displacement field ${\bf u}$
$$
u_x=ARe \left[\frac{(x+i\alpha_dy)^{\lambda}}{i\alpha_d}+
i\alpha_s(x+i\alpha_sy)^{\lambda}\right],
$$
\be
u_y=ARe \left[(x+i\alpha_dy)^{\lambda}-(x+i\alpha_sy)^{\lambda}\right]
\ee
with
$\alpha_d^2=1-(V_{tip}/C_d)^2$ and 
$\alpha_s^2=1-(V_{tip}/C_s)^2$,  where $C_d$ and $C_s$ are dilation and
shear wave speed. The spectrum of $\lambda$ is still given by Eq. (6)
but now $\varepsilon$ reads
\be
\varepsilon= \frac{2}{\pi}\frac{(1-\alpha_d\alpha_s)}
{\alpha_d(1-\alpha_s^2)} \frac{V_{tip}}{S}.
\ee 
Thus, elastodynamic effects just lead  to a redefinition of 
$\varepsilon $ in the main
result, Eq. (11), which remains valid. 
For small velocities, Eqs. (14) and (15) reduce to Eqs.(5) and 
(7), respectively.

The most serious problem with  large  velocities 
arises in connection 
with   a self-consistent description of the dissipation at
the tip. This part of  kinetics cannot be considered
macroscopically for arbitrary tip velocities. One can only treat 
the case of small velocities, $V_{tip}\ll C_s$ in a model
independent way, by introducing the tip kinetic coefficient. For
higher velocities the so-called velocity dependent fracture energy
$\gamma (V_{tip})$ is introduced. This function contains
information about the usual surface energy $\gamma$ and tip
dissipation and reduces to the surface energy in the static limit.
The main dissipation in our approach arises from the friction
between both sides of the crack and can be treated
macroscopically. Note that this part of the dissipation can be
described by the velocity independent  friction
coefficient $\alpha$ even at  large tip velocities in the case of
small sliding velocities.

Up to now we have discussed  shear cracks along the interface
between two dissimilar materials (the case of a rigid substrate).
We also present the result for the case of frictional shear cracks
inside homogeneous elastic materials. Such cracks can propagate under the
 shear loading in amorphous materials, or along grain boundaries in
crystals (in single crystals the  well known dislocation mechanism of
plasticity should be favorable).  The boundary conditions
on the crack surfaces, which remain in contact,
are: continuous normal displacement and 
continuous normal and shear stresses. The shear stress is also
given by Eq.(1) with $\dot u_x$ being the relative sliding
velocity of two crack surfaces. We note that these boundary
conditions are quite different from the standard boundary
conditions of mode II (and III) cracks: zero normal and shear
stresses on the crack surfaces \cite{Freund}. In our case, the
frictional shear crack, we find the following expression for
$\varepsilon $ which enters the general result, Eq.(11):
\be
\varepsilon= \frac{4}{\pi}\frac{\alpha_s(1-\alpha_s^2)}{4\alpha_d\alpha_s-
(1+\alpha_s^2)^2}
\frac{V_{tip}}{ S}.
\ee

Now let us discuss the relation of our results to  experimental
observations of Baumberger, Caroli and Ronsin
\cite{Baumberger2002}. They performed  experiments of a gel
sliding on a glass plate. The driving velocity was given and the
shear force and thus the average shear stress was deduced from the
spring elongation. Above some critical driving velocity
$V_c\approx 125\mu m/s$, steady sliding was observed. At 
velocities smaller than the critical one, periodic stick slip sets up
(see figures in \cite{Baumberger2002}). Upon increasing the
driving velocity $V$, no hysteresis of the transition was
detected. In the stick slip regime they observed the propagation
of self-healing pulses with no opening,  nucleated periodically at
the trailing edge of the sample. The propagation velocity of these
cracks was about 60 times larger than the critical sliding
velocity, yet still much smaller than the  shear wave speed.

The existence of a critical sliding velocity, where stationary
sliding is stable against the healing instability, appears
naturally in our description and is given by Eq. (3). The
characteristic value of the shear strain in the sliding mode near
the critical velocity experimentally was about  $u_{xy}= 0.04$. 
Thus, we can
estimate from Eq. (2)  the characteristic velocity $S=1.5mm/s$
and from Eq.(3) we find that the characteristic difference between
the interface energies in the slip and stick states is
$\gamma=0.1J/m^2$. One would expect that for ordinary elastic
materials the velocity $S$ should be of the order of the speed of sound.
However, for gels the shear modulus $\mu$ is much smaller
than for ordinary materials. The shear wave speed
$C_s=(\mu/\rho)^{1/2}$ is only $2m/s$. The velocity
$S=\mu/\alpha$ is linear in $\mu$ and should be even smaller.
This is a possible explanation
for a relatively small value of $S$ compared to $C_s$.

In the stick-slip regime, which exists below $V_c$, the nucleation of
a slip pulse takes place at the trailing edge of the sample and
requires overshooting above the Griffith threshold according to
the experiment. This overshooting is not small and in order to estimate
the crack velocity we can use Eq. (12) since the velocity is still much
smaller than the  speed of sound. Because of the week logarithmic
parameter dependence, we conclude that $V_{tip}$ is of the order
of $S$ and essentially independent of the driving velocity in agreement
with experimental observations.
After the slip pulse traversed the sample the stress drops
below the Griffith threshold and resticking takes place via propagation
of a healing pulse. Its velocity is still described by Eq. (12) with
$\Delta <1$. This periodic stick slip regime bifurcates towards
stationary sliding at $V=V_c$ where  stresses are always above the
Griffith threshold. For driving velocities slightly below $V_c$, 
characteristic values of $\Delta$ for resticking 
are close to $1$. 
Since the velocity of the resticking crack is small and  
comparable with the driving velocity in this range, 
a complicated collective behavior of self-healing pulses is observed 
\cite{Baumberger2002}.

While our results are in a qualitative agreement with experimental
observations, we still underestimate the crack velocity which is a
few times smaller than in the experiment. The other discrepancy is
due to the observed nonlinear behavior of the stress with the
velocity in the stationary sliding regime 
 (the so-called  shear-thinning rheology). On the other
hand, the geometry of the experiment is such that the total
macroscopic friction of the sliding sample obviously depends on
the processes  taking place at the edges of the sample. The
stresses here are highly inhomogeneous and the kinetic phenomena
should be considered with a great care.  
We think that
these boundary effects may be responsible for the 
discrepancies mentioned above.

Another subject of future investigations should be the collective
behavior of self-healing pulses in the spirit of Ref. \cite{Caroli2000}.
Further theoretical and experimental investigations are needed to
shed light on this phenomenon where two intriguing problems, crack
propagation and friction, combine together.

Discussions with S.V. Iordanskii, B.N.J. Persson and D.E. Temkin
are greatly appreciated. V.M. thanks Forschungszentrum J\"ulich
for its hospitality.

\end{multicols}

\begin{thebibliography}{999}
\bibitem{Rubio94}
M. Rubio and J. Galeano, Phys. Rev. E {\bf 50}, 1000 (1994).
\bibitem{Baumberger2002}
T. Baumberger, C. Caroli, and O. Ronsin, Phys. Rev. Lett. 88,
075509 (2002).
\bibitem{Persson2000}
B. N. J. Persson, {\it Sliding Friction: physical principles and 
applications} (Springer, Heidelberg, 2000).
\bibitem{Gerde01}
E. Gerde and M. Marder, Nature (London), {\bf 413}, 285 (2001).
\bibitem{Heaton90}
T. Heaton, Phys. Earth Planet. Inter.  {\bf 64}, 1 (1990).
\bibitem{Ranjith01}
K. Ranjith and J. Rice, J. Mech. Phys. Solids {\bf 49}, 341 (2001).
\bibitem{Caroli2000}
C. Caroli, Phys. Rev. E {\bf 62}, 1729 (2000).
\bibitem{Freund}
L.B. Freund, {\it Dynamic Fracture Mechanics} (Cambridge
University Press, New York, 1990).
\end{thebibliography}
\end{document}